\documentclass[12pt]{iopart}

\usepackage{graphics}
\begin{document}

\title[Indication for $\pi^+ \pi^-$ scattering in $p+p$ collisions at $\sqrt{s_{_{NN}}} = $
200 GeV]{Evidence for $\pi^+ \pi^-$ scattering in $p+p$ collisions
at $\sqrt{s_{_{NN}}} = $ 200 GeV}

\author{P. Fachini, R. S. Longacre, Z. Xu and H. Zhang}

\address{Brookhaven National Laboratory, Upton, NY, 11973, USA}
\ead{pfachini@bnl.gov}
\begin{abstract}
 A $\rho(770)^0$ mass shift of about -40 MeV/$c^2$ was
measured in $p+p$ collisions at $\sqrt{s_{_{NN}}} = $ 200 GeV at
RHIC. Previous mass shifts have been observed at CERN-LEBC-EHS and
CERN-LEP. We will show that phase space does not account for the
$\rho(770)^0$ mass shift measured at RHIC, CERN-LEBC-EHS and
CERN-LEP and conclude that there are significant scattering
interactions in $p+p$ collisions.
\end{abstract}

\maketitle

\section{Introduction}
\label{sec:intro}

The $\rho^0$ was measured via its hadronic decay channel in
minimum bias $p+p$ and Au+Au collisions at RHIC and a mass shift
of -40 and -70 MeV/$c^2$ of the position of the $\rho^0$ was
observed, respectively. While no explanations were explicitly
attributed to the mass shift in $p+p$ collisions, the possible
explanations for the apparent modification of the $\rho^0$ meson
properties in Au+Au collisions were attributed to dynamical
interactions with the surrounding matter, interference between
various $\pi^+\pi^-$ scattering channels, phase space distortions
due to the rescattering of pions forming $\rho^0$ and
Bose-Einstein correlations between $\rho^0$ decay daughters and
pions in the surrounding matter \cite{rho}.

The $\rho^0$ meson measured in the dilepton channel probes all
stages of the system formed in relativistic heavy-ion collisions
because the dileptons have negligible final state interactions
with the hadronic environment. Heavy-ion experiments at CERN show
an enhanced dilepton production cross section in the invariant
mass range of 200-600 MeV$/c^2$, showing that the $\rho^0$ is
broadened rather than shifted \cite{aga,na60}. The hadronic decay
measurement at RHIC \cite{rho}, $\rho(770)^0 \rightarrow
\pi^{+}\pi^{-}$, was the first of its kind in heavy-ion
collisions. Since the $\rho^0$ lifetime of $c\tau$ = 1.3 fm is
small with respect to the lifetime of the system formed in Au+Au
collisions, the $\rho^0$ meson is expected to decay, regenerate,
and rescatter all the way through kinetic freeze-out. Therefore,
the measured $\rho^0$ mass at RHIC should reflect conditions at
the late stages of the collisions \cite{brown,rapp1}. However, its
has been shown that the $\rho^0$ width goes to zero with the
$\rho^0$ mass dropping near the chiral phase transition, which is
called the the vector manifestation and it can occur near $T_c$ in
the hot and dense matter \cite{harada}. In this scenario, the
$\rho^0$ lifetime becomes very large and the $\rho^0$ produced at
the early stages can also be measured in the hadronic channel.

The modification of the $\rho^0$ properties in heavy-ion
collisions has been expected \cite{rapp}, contrary to the
modifications of the $\rho^0$ properties in $p+p$ collisions.
Similar modifications of that measured in $p+p$ collisions at RHIC
has been observed before at CERN-LEBC-EHS and CERN-LEP. Until now,
the mass shift measured in $p+p$ at CERN-LEBC-EHS has been
attributed to phase space. The process of hadron-hadron collisions
at low $p_T$ is not well understood. These collisions are viewed
as a complicated interaction, where a perturbative treatment of
the interaction dynamics is not possible. As a consequence, only
phenomenological models with assumptions about the partonic
subprocess dynamics can be used to interpret the data. By studying
soft hadronic interactions we may be able to explain the
interaction dynamics of hadron collisions and possibly understand
the parton hadronization mechanism, which is important in jet
studies. One way of probing the dynamics of hadron-hadron
collisions is to study the production of resonances, in
particular, the $\rho^0$ meson.

In this paper, we will show that phase space does not account for
the $\rho(770)^0$ mass shift measured at RHIC \cite{rho},
CERN-LEBC-EHS \cite{na27} and CERN-LEP \cite{act,laf,bus,ack} and
conclude that there are significant scattering interactions in
$p+p$ collisions.

\section{Discussion $\rho$ mass average from the Particle Data Group (PDG)}
\label{sec:pdg}

We will first discuss the $\rho$ mass average from the PDG
\cite{pdg}. The $\rho^0$ mass average 775.8 $\pm$ 0.5 MeV/$c^2$
from $e^+e^-$ was obtained from either $e^+e^- \rightarrow
\pi^+\pi^-$ or $e^+e^- \rightarrow \pi^+\pi^-\pi^0$. This means
that the $\rho^0$ mass average was obtained from {\it exclusive
leptonic reactions}. Similarly, the $\rho^{\pm}$ mass average
775.5 $\pm$ 0.5 MeV/$c^2$ was also was obtained from {\it
exclusive leptonic reactions}.

The $\rho$ averages reported by the PDG from reactions other than
leptonic interactions are systematic lower than the value obtained
from {\it leptonic exclusive interactions} by $\sim$ 10 MeV/$c^2$
\cite{pdg}. The $\rho$ production in these hadronic reactions are
inclusive and exclusive. In the case of inclusive productions, the
phase space was take into account when the $\rho$ mass was
measured (e.g. \cite{na27}).

These observations lead us to conclude that the $\rho$ mass
depends on specific interactions, e.g. whether the $\rho$ is
produced in inclusive or exclusive reactions. Since a leptonic
reaction and exclusive measurement of the $\rho^0$ lead to a
negligible modification of any kind of the $\rho^0$ mass, the
average 775.8 $\pm$ 0.5 MeV/$c^2$ \cite{pdg} from $e^+e^-$ should
correspond to the $\rho^0$ mass in the vacuum.

\section{$\rho^0$ mass shifts at RHIC, CERN-LEBC-EHS, and CERN-LEP}
\label{sec:mass}

\begin{figure}[tbp]
\begin{center}
\resizebox{0.47\textwidth}{!}{%
  \includegraphics{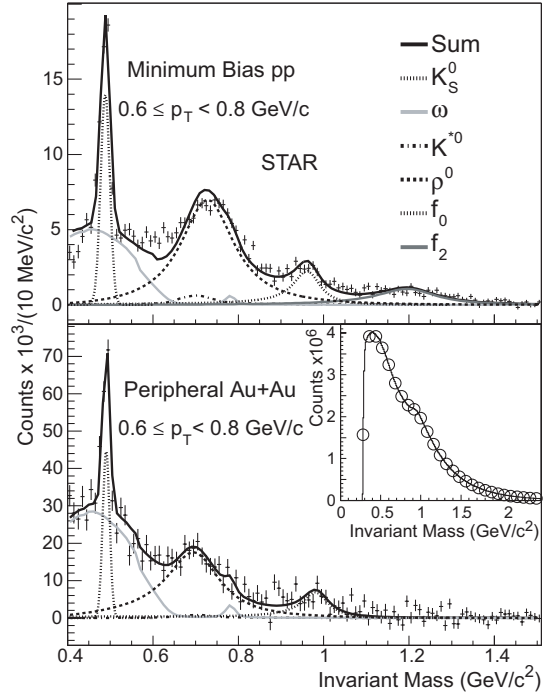}
  }
\caption{The raw $\pi^+\pi^-$ invariant mass distributions after
subtraction of the like-sign reference distribution for minimum
bias $p+p$ (top) and peripheral Au+Au (bottom) interactions
measured by STAR. For details see \cite{rho}}\label{fig:cocktail}
\end{center}
\end{figure}

\begin{figure}[tbp]
\begin{center}
\resizebox{0.47\textwidth}{!}{%
  \includegraphics{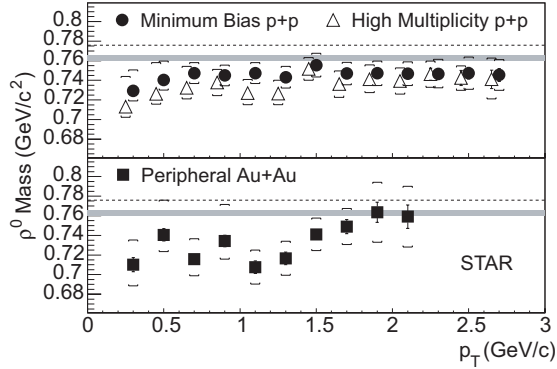}
  }
\caption{The $\rho^0$ mass as a function of $p_T$ for minimum bias
$p+p$ (filled circles), high multiplicity $p+p$ (open triangles),
and peripheral Au+Au (filled squares) collisions measured by STAR.
The error bars indicate the systematic uncertainty. Statistical
errors are negligible. The $\rho^0$ mass was obtained by fitting
the data to the BW$\times$PS functional form described in
\cite{rho}. The dashed lines represent the average of the $\rho^0$
mass measured in $e^+e^-$ \cite{pdg}. The shaded areas indicate
the $\rho^0$ mass measured in $p+p$ collisions \cite{na27}. The
open triangles have been shifted downward on the abscissa by 50
MeV/$c$ for clarity.}\label{fig:mass}
\end{center}
\end{figure}

The $\rho^0$ was measured in the hadronic decay channel $\rho^0
\rightarrow \pi^{+}\pi^{-}$ at RHIC, CERN-LEBC-EHS, and CERN-LEP
in inclusive production. At RHIC, the STAR collaboration measured
the $\rho^0$ at $\sqrt{s_{_{NN}}} = $ 200 GeV at midrapidity ($|y|
< 0.5$) and observed mass shifts of the position of the $\rho^0$
peak of about -40 MeV/$c^2$ and -70 MeV/$c^2$ in minimum bias
$p+p$ and peripheral Au+Au collisions, respectively \cite{rho}.
The invariant mass distributions from \cite{rho} is shown in Fig.
\ref{fig:cocktail}. The solid black line in Fig.
\ref{fig:cocktail} is the sum of all the contributions in the
hadronic cocktail. The $K_S^0$ was fit to a Gaussian (dotted
line). The $\omega$ (light grey line) and $K^{\ast}(892)^{0}$
(dash-dotted line) shapes were obtained from the HIJING event
generator \cite{hijing}, with the kaon being misidentified as a
pion in the case of the $K^{\ast 0}$. The $\rho^0(770)$ (dashed
line), the $f_0(980)$ (dotted line) and the $f_2(1270)$ (dark grey
line) were fit by relativistic Breit-Wigner functions (BW) times
the Boltzmann factor (PS) to account for phase space
\cite{brown,barz,pbm,kolb,rapp2,bron,prat,gran},
\begin{equation}
\mathrm{BW} = \frac{M_{\pi\pi}M_0\Gamma}{(M_0^2 - M_{\pi\pi}^2)^2
+ M_0^2\Gamma^2} \label{bwstar}
\end{equation}
\begin{equation}
\Gamma = \Gamma_0 \times
\frac{M_0}{M_{\pi\pi}}\times[\frac{M_{\pi\pi}^2 - 4m_\pi^2}{M_0^2
- 4m_\pi^2}]^{(2\ell+1)/2}
\end{equation}
\begin{equation}
\mathrm{PS} = \frac{M_{\pi\pi}}{\sqrt{M_{\pi\pi}^2 + p_T^2}}
\times \exp\frac{-\sqrt{M_{\pi\pi}^2 + p_T^2}}{T}
\end{equation}
where $M_0$ and $\Gamma_0$ are the natural resonance mass and
width, respectively \cite{rho,pdg}. The masses of $K_S^0$,
$\rho^0$, $f_0$, and $f_2$ were free parameters in the fit, and
the widths of $\rho^0$, $f_0$ and $f_2$ were fixed according to
\cite{pdg}. The uncorrected yields of $K_S^0$, $\rho^0$, $\omega$,
$f_0$, and $f_2$ were free parameters in the fit while the
$K^{\ast 0}$ fraction was fixed according to the
$K^{\ast}(892)^{0} \!\rightarrow\! \pi K$ measurement
\cite{kstar}. The $\rho^0$, $\omega$, $K^{\ast 0}$, $f_0$, and
$f_2$ distributions were corrected for the detector acceptance and
efficiency determined from a detailed simulation of the TPC
response using GEANT \cite{rho}. The number of degrees of freedom
(d.o.f.) from the fits was 196 and the typical $\chi^2/$d.o.f. was
1.4.
The $\rho^0$ mass obtained from the BW$\times$PS fit is depicted
in Fig. \ref{fig:mass}, where it is clear that the phase space
does not account for the measured mass shifts of the position of
the $\rho^0$ peak.

\begin{figure}[tbp]
\begin{center}
\resizebox{0.48\textwidth}{!}{%
  \includegraphics{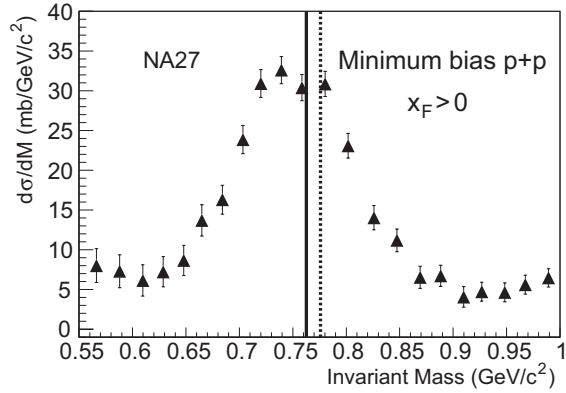}
  }
\caption{The invariant $\pi^+\pi^-$ mass distribution after
background subtraction (equation \ref{background}) for minimum
bias $p+p$ collisions measured by NA27. For details see
\cite{na27}. The vertical dashed line represent the average of the
$\rho^0$ mass 775.8 $\pm$ 0.5 MeV/$c^2$ measured in $e^+e^-$
\cite{pdg}. The vertical solid line is the $\rho^0$ mass 762.6
$\pm$ 2.6 MeV/$c^2$ reported by NA27 \cite{na27}.
}\label{fig:massNA27}
\end{center}
\end{figure}

\begin{figure}[tbp]
\begin{center}
\resizebox{0.48\textwidth}{!}{%
  \includegraphics{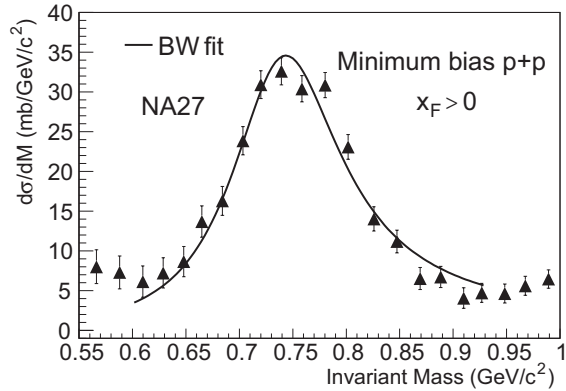}
  }
\caption{The invariant $\pi^+\pi^-$ mass distribution after
background subtraction (equation \ref{background}) for minimum
bias $p+p$ collisions measured by NA27 \cite{na27} fit to the BW
function (equation \ref{bwstar}). For details see \cite{na27}.
}\label{fig:fit}
\end{center}
\end{figure}

At CERN-LEBC-EHS, NA27 measured the $\rho^0$ in minimum bias $p+p$
at $\sqrt{s} = $ 27.5 GeV for $x_F > 0$, where $x_F$ is the ratio
between the longitudinal momentum and the maximum momentum of the
meson, and reported a mass of 762.6 $\pm$ 2.6 MeV/$c^2$
\cite{na27}. The $\rho^0$ signal can be seen even before
background subtraction. The invariant $\pi^+\pi^-$ mass
distribution was fit to the BW(NA27)$\times$PS(NA27) plus a
background function \cite{na27}
\begin{eqnarray}
\mathrm{BW(NA27)} = \frac{2M_{\pi\pi}}{(M_{\pi\pi}^2 -
4m_\pi^2)^{1/2}} \times \nonumber \\
\frac{\Gamma}{(M_0^2 - M_{\pi\pi}^2)^2 + M_0^2\Gamma^2}
\label{bwna27}
\end{eqnarray}
\begin{equation}
\Gamma = \Gamma_0 \times
\frac{M_0}{M_{\pi\pi}}\times[\frac{M_{\pi\pi}^2 - 4m_\pi^2}{M_0^2
- 4m_\pi^2}]^{(2\ell+1)/2}
\end{equation}
\begin{eqnarray}
\mathrm{PS(NA27) = BG} = \frac{\alpha_1}{(M_{\pi\pi})^{\alpha_2}}
[\frac{M_{\pi\pi}^2 - 4m_\pi^2}{M_{\pi\pi}}]^{\alpha_2/2} \times
\nonumber \\
 \exp(-\alpha_3[\frac{M_{\pi\pi}^2 -4m_\pi^2}{M_{\pi\pi}}]^{1/2}
 -\alpha_4[\frac{M_{\pi\pi}^2 - 4m_\pi^2}{M_{\pi\pi}}])
 \label{background}
\end{eqnarray}
where $\alpha_1, \alpha_2, \alpha_3,$ and $\alpha_4$ are free
parameters in the fit. In this analysis, the phase space function
(PS(NA27)) used is the same as the combinatorial background (BG).
The invariant $\pi^+\pi^-$ mass distribution after subtraction of
the mixed-event reference distribution is shown in Fig.
\ref{fig:massNA27}. The vertical dashed line represent the average
of the $\rho^0$ mass 775.8 $\pm$ 0.5 MeV/$c^2$ measured in
$e^+e^-$ \cite{pdg}. The vertical solid line is the $\rho^0$ mass
762.6 $\pm$ 2.6 MeV/$c^2$ reported by NA27 \cite{na27}. As shown
in Fig. \ref{fig:massNA27}, the position of the $\rho^0$ peak is
shifted by $\sim$ 30 MeV/$c^2$ compared to the $\rho^0$ mass in
the vacuum 775.8 $\pm$ 0.5 MeV/$c^2$ \cite{pdg}. The shift of the
$\rho^0$ peak can be quantified by fitting the NA27 $\pi^+\pi^-$
mass distribution after subtraction of the mixed-event reference
distribution to the BW function (equation \ref{bwstar}). The fit
is depicted in Fig. \ref{fig:fit} and the value of the maximum of
the distribution obtained from the fit is 747.6 $\pm$ 2.0
MeV/$c^2$ with $\chi^2/ndf$ = 1.9. If equation \ref{bwna27} is
used instead, the value of the maximum of the distribution is
754.3 $\pm$ 2.1 MeV/$c^2$ with $\chi^2/ndf$ = 1.2. The difference
of a few MeV between the two values is due to the different BW
functions used in the analyzes.

As mentioned, NA27 obtained the $\rho^0$ mass by fitting the
invariant $\pi^+\pi^-$ mass distribution to the BW$\times$PS
function, and they reported a mass of 762.6 $\pm$ 2.6 MeV/$c^2$,
which is $\sim$ 10 MeV/$c^2$ lower than the $\rho^0$ mass in the
vacuum. Ideally, the PS factor should have accounted for the shift
on the $\rho^0$ peak, and the mass obtained from the fit should
have agreed with the $\rho^0$ mass in the vacuum. However, just
like in the STAR measurement, this was not the case, since the
phase space did not account for the mass shift on the position of
the $\rho^0$ peak.

At CERN-LEP, OPAL, ALEPH and DELPHI measured the $\rho^0$ in
inclusive $e^+e^-$ reactions at $\sqrt{s} = $ 90 GeV
\cite{act,laf,ack,bus}. Even though OPAL never reported the value
of the $\rho^0$ mass, OPAL reported a shift on the position of the
$\rho^0$ peak by $\sim$ 70 MeV/$c^2$ at low $x_p$, where $x_p$ is
the ratio between the meson and the beam energies, and no shift at
high $x_p$ ($x_p \sim 1$) \cite{act,laf}. OPAL also reported a
shift in the position of the $\rho^{\pm}$ peak from -10 to -30
MeV/$c^2$, which was consistent with the $\rho^0$ measurement
\cite{ack}. ALEPH reported the same shift on the position of
$\rho^0$ peak observed by OPAL \cite{bus}. Both OPAL and ALEPH
used the like-sign method to subtract the background. DELPHI fit
the raw invariant $\pi^+\pi^-$ mass distribution to the
(BW$\times$PS + BG) function
\begin{eqnarray}
\mathrm{PS(DELPHI) = BG} = ( M - M_{th})^{\gamma_1} \times
\nonumber
\\
\exp(\gamma_2M + \gamma_3M^2 + \gamma_4M^3 + \gamma_5M^4),
\end{eqnarray}
where $M_{th}$ is the threshold mass and BW is a relativistic
Breit-Wigner function, for $x_p >$ 0.05 and reported a $\rho^0$
mass of 757 $\pm$ 2 MeV/$c^2$ \cite{abr}, which is five standard
deviations below the $\rho^0$ mass in the vacuum (775.8 $\pm$ 0.5
MeV/$c^2$). As one can see, similarly to NA27, DELPHI assumed that
the phase space was described by the background function.
Bose-Einstein correlations were used to describe the shift on the
position of $\rho^0$ peak. However; high (even unphysical)
chaocity parameters ($\lambda \sim 2.5$) were needed
\cite{act,laf,bus}.

\section{Phase space in $p+p$ collisions}
\label{sec:phasespacepp}

\begin{figure}[tbp]
\begin{center}
\resizebox{0.47\textwidth}{!}{%
  \includegraphics{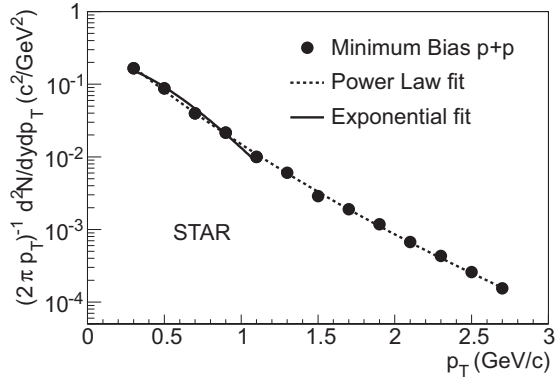}
  }
\caption{The $p_T$ distribution at $|y| \!<\!$ 0.5 for minimum
bias $p+p$ collisions \cite{rho}. The black circles are the data,
the dashed line is the power-law fit and the solid line is the
exponential fit. See \cite{rho} for the fit functions. The errors
shown are statistical only and smaller than the
symbols.}\label{fig:spectra}
\end{center}
\end{figure}

In $p+p$ collisions, most models assume that particles are born at
hadronization according to phase space without any final state
interaction. In multiparticle production processes,
single-inclusive (e.g. $p+p$), invariant particle spectra are
typically exponential in $p_T$ \cite{hage}. The exponential
behavior does not require final state interactions and it can be
due to phase space population at hadronization. The slope
parameter in $p+p$ collisions are independent of the particle
species \cite{bear}. At RHIC, the $\rho^0$ spectra in minimum bias
and high multiplicity $p+p$ are exponential in $p_T$ up to 1.1
GeV/$c^2$ with slope parameters of $\sim$ 180 MeV. The $\rho^0$
spectrum in minimum bias $p+p$ \cite{rho} is depicted in Fig.
\ref{fig:spectra}. For reference, the slope parameter of $\pi^-$
is $\sim$ 160 MeV \cite{pion}. Note that the slope parameter in
$p+p$ is independent of the particle rest mass and shows $m_T$
scaling. These results are also independent of the multiplicity.

However, the $\rho^0$ mass measured in $p+p$ at RHIC is
multiplicity dependent (the mass shift in high multiplicity is
higher than in minimum bias $p+p$ collisions) \cite{rho}, which is
opposite to the slope parameter that is multiplicity independent.
Note that the systematic errors are correlate among themselves and
between the two measurements. We will demonstrate that the mass
should be independent of $p_T$ as well as multiplicity at
hadronization without final state interactions. According to
quantum mechanics, a resonance at rest is described by the wave
function,
\begin{equation}
\Psi(x,t) \propto \exp(\frac{-iE_0t}{\hbar}) \times
\exp(\frac{-t}{2\tau})
\end{equation}
where $\tau$ is the lifetime and $E_0$ is the energy at rest. The
probability amplitude of the resonance decay can be written as
\begin{equation}
\tilde{\Psi}(E,x) \propto \int_0^\infty
(\exp(\frac{-iEt}{\hbar}))^* \Psi(x,t) dt
\end{equation}
\begin{equation}
\tilde{\Psi}(E,x) \propto
\frac{1}{i\frac{E-E_0}{\hbar}-\frac{1}{2\tau}} \Psi(x)
\end{equation}
\begin{equation}
P(E) \propto |\frac{1}{i\frac{E-E_0}{\hbar}-\frac{1}{2\tau}}|^2
\int|\Psi(x)|^2dx
\end{equation}
\begin{equation}
P(E) \propto \frac{\Gamma/2}{(E-E_0)^2+\frac{\Gamma^2}{4}}
\label{bw}
\end{equation}
where $\Gamma = \hbar/\tau$, and the probability amplitude is a
non-relativistic Breit-Wigner distribution. When the energy
conservation law is imposed in the partition of string
fragmentation into multiple hadrons, a phase space factor, similar
to the Boltzmann factor in thermal distribution, has to be added
to equation \ref{bw}. Replacing $E$ by the invariant mass ($M$),
the phase space is $\exp(\frac{-m_T}{T})$, where $m_T$ equals
$\sqrt{M^2+p_T^2}$ and T equals 160 MeV. Equation \ref{bw} is then
rewritten as
\begin{equation}
P(M,p_T) \propto \frac{\Gamma/2}{(M-M_0)^2+\frac{\Gamma^2}{4}}
\times \exp(\frac{-m_T}{T}). \label{bwps}
\end{equation}
However; as discussed previously, the phase space does not
describe the mass shift of the $\rho$ meson measured at RHIC,
CERN, and LEP.

Most event generators (e.g. PYTHIA \cite{sjo} and HIJING
\cite{wan}) create resonances according to a non-relativistic
Breit-Wigner function at a given $p_T$ (equation \ref{bw}).
However, we just showed that a more reasonable way to produce
resonances is to use equation \ref{bwps}.

Since the phase space of a non interacting multiparticle state
{\it cannot} explain the distortion of the $\rho^0$ line shape, we
can conclude that the phase space in $p+p$ collisions also
accounts for hadrons scattering and forming resonances. In the
case of the $\rho^0$, $\pi^+\pi^- \rightarrow \rho^0 \rightarrow
\pi^+\pi^-$. This can be pictured in the string fragmentation
particle production scenario, where the string breaks several
times, two pions are formed, they scatter, and form a $\rho^0$.
Such interactions are significant and modify the $\rho^0$ spectral
shape in $e^+e^-$ and $p+p$ from a relativistic $p$-wave
Breit-Wigner function.

Using the $p+p$ data from the top plot of Fig. \ref{fig:cocktail}
and the rescattering formalism of \cite{longacre}, we can test the
ideas of a mass shift having a $\pi^+\pi^-$ scattering component.
The di-pion production is given by equation 21 of \cite{longacre}.
For the $\rho^0$ which is p-wave $\ell=1$ equation 21 becomes
\begin{equation}
|T|^2  = |D|^2 \frac{sin^2(\delta_{1})}{\mathrm{PS'}} +
\frac{|A|^2}{\mathrm{PS'}} \left| \alpha sin(\delta_1) +
\mathrm{PS'} cos(\delta_1) \right|^2 \label{fit}
\end{equation}
The $\delta_{1}$ is the p-wave phase shift ($\rho^0$) and
$\frac{sin^2(\delta_{1})}{\mathrm{PS'}}$ in our case becomes the
Breit-Wigner times phase space equation \ref{bwps}. $|D|^2$
becomes the direct production of $\rho^0$ for the $p_T$ range 600
to 800 MeV/c. $|A|^2$ is the phase space overlap of di-pions in
the $\ell =$ 1 partial waves. The pions emerge from a close
encounter in a defined quantum state with a random phase. The
emerging pions can re-interact or re-scatter through the p-wave
quantum state or a phase shift. The phase space overlap comes from
sampling the $\pi$ spectrum from $p+p$ collisions, where the sum
of the $\pi^+\pi^-$ has the correct $p_T$ range. The di-pion mass
spectrum for this $p_T$ range falls with increasing di-pion
invariant mass. The variable $\alpha$ is related to the real part
of the di-pion rescattering diagram and measures the range of
interaction \cite{longacre}
\begin{equation}
\alpha = (1.0 - \frac{r^2}{r_0^2})
\end{equation}
The $r$ is the radius of rescattering in fm and $r_0$ is 1.0 fm or
the limiting range of the strong interaction. The PS' is the phase
space factor
\begin{equation}
\mathrm{PS'} =
\frac{2q(q/q_0)^2}{\frac{M(q/q_0)^2}{(1+(q/q_0)^2)}}
\end{equation}
where $q$ is the momentum of the pions in the center-of-mass and
$q_0$ is the size of interaction for the $\rho^0$ system and we
use 1.0 fm distance to set $q_0$ at 200 MeV/$c$.

\begin{figure}[tbp]
\begin{center}
\resizebox{0.47\textwidth}{!}{%
  \includegraphics{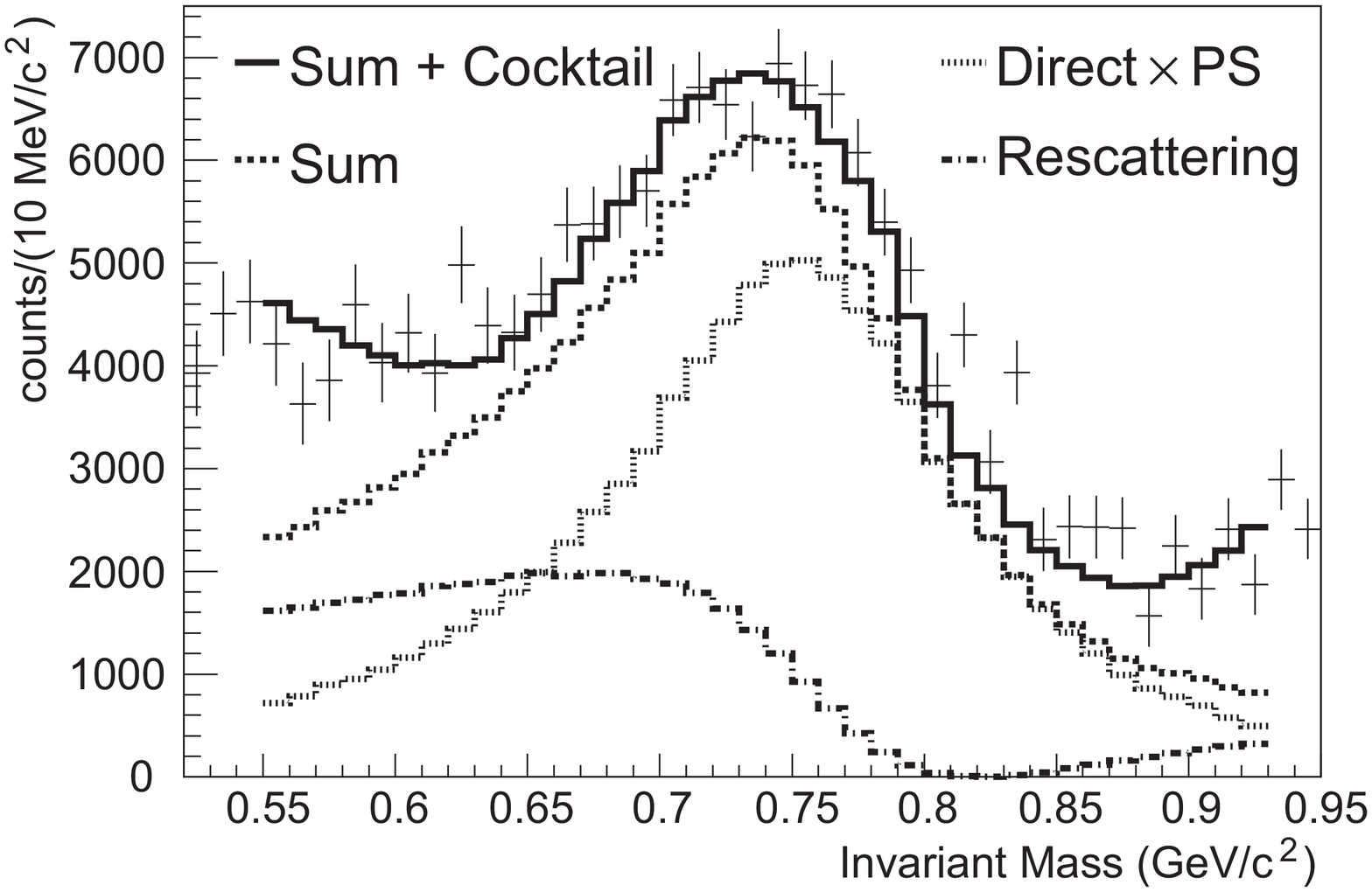}
  }
\caption{Fit to the data using equation \ref{fit} (solid line),
the direct $\rho^0$ contribution term times the phase space
(dashed line), the rescattering term (dash-dotted line) and the
sum of the both (dashed line). }\label{fig:ron}
\end{center}
\end{figure}

In order the fit the $p+p$ data we need to add the rest of the
cocktail of resonances used by STAR \cite{rho}. Such a fit is
shown in Fig. \ref{fig:ron} (solid line), where the direct
$\rho^0$ term time the phase space (dotted line) and the
rescattered term (dash-dotted line) plus the sum (dashed line) are
also depicted . In equation \ref{fit} we have three parameters
which are fitted. A scale factor on the phase space overlap of the
di-pions. The direct production $D$ and $\alpha$. The value for
$\alpha$ form the fit is 0.47, which implies $r$ = 0.73 fm.

We see that the mass shift is a rescattering effect coming from a
$\pi \pi$ source which is not the $\rho$. This $\pi \pi$ source
becomes a $\rho^0$ through reinteraction. The mass shift of the
$\rho$ seen in \cite{UPC}, where the $\rho^0$ is photoproduced, is
due to the same rescattering effect, except that the direct
production of the $\rho^0$ in this case is coherent with the $\pi
\pi$ source.

STAR measured the $\rho^0$ at midrapidity and NA27 measured at the
forward region. In addition, there is the energy difference of
$\sqrt{s_{_{NN}}} = $ 200 and 27.5 GeV, respectively. All these
facts are consistent our model, the difference between the
$\rho^0$ mass shift measure by STAR and NA27 is due to pion
multiplicity. Similar arguments is valid for the $\rho$
measurements at CERN-LEP at $\sqrt{s_{_{NN}}} = $ 90 GeV
indicating that there is also significant scattering interactions
in $e^+e^-$ reactions.

\section{Conclusions}
\label{sec:conclusions}

We discussed that the natural mass of the $\rho$ meson should be
measured in {\it exclusive} reactions only. We showed that a shift
on the position of the $\rho^0$ peak has been measured before and
that the phase space does not account for the $\rho(770)^0$ mass
shift measured at RHIC, CERN-LEBC-EHS, and CERN-LEP. In addition,
we discussed the phase space in $p+p$ collisions and concluded
that there are significant scattering interactions in $p+p$
reactions. These interactions modify the $\rho^0$ line shape in
$p+p$ and $e^+e^-$ interactions from a Breit-Wigner function.
Using the rescattering formalism of \cite{longacre}, we can
reproduce the $\rho^0$ mass shift measured by the STAR
collaboration in minimum bias $p+p$ collisions.

\section{Acknowledgement}
\label{sec:acknowledgement}

This work was supported in part by the HENP Divisions of the
Office of Science of the U.S. DOE.

\end{document}